\documentclass[12pt]{article}

\global\arraycolsep=1pt
\usepackage{amssymb}
\usepackage{amsmath}
\usepackage {pstricks}
\usepackage{graphicx,color}
\usepackage{mathrsfs}

\usepackage{caption} 
\renewcommand{\thefootnote}{\fnsymbol{footnote}}

\definecolor{usuzumiiro}{cmyk}{0.08,0.05,0.06,0.55}

\numberwithin{equation}{section}
\begin{document}

%
\begin{titlepage}
\begin{flushright}
{RIKEN-MP-79}\\
\end{flushright}
\vspace{0.5cm}
\begin{center}
{\Large \bf 
\textsl{}
\textsc{
Notes on Enhancement of Flavor Symmetry \vspace{0.3cm}
\\
and \vspace{0.3cm}
5d Superconformal Index
}}
\vskip1.0cm
{\large Masato Taki\footnote[2]{Email: taki@riken.jp}
}
\vskip 1.0em
{\it
Mathematical Physics Lab., RIKEN Nishina Center,\\ 
Saitama 351-0198, Japan
}%
\end{center}
\vskip1.5cm

\begin{abstract}
The UV fixed point theory of $SU(2)$ gauge theory with $N_f=0,1,\cdots,7$ flavors
 is believed to have the enlarged $E_{N_f+1}$ flavor symmetry.
Actually it is not easy to check this conjecture because the UV theory is strongly-coupled,
however, computation of certain SUSY protected quantities provides strong evidence 
for the enhancement of flavor symmetry.
We study the superconformal index
 for $SU(2)$ gauge theory with $N_f=0, 1$ flavors in details,
and we give a support for
the enhancement 
by studying combinatorial structure of the superconformal indexes of these theories.
We also give a nontrivial evidence that the local $\mathbb{F}_2$ geometry leads to 
the $E_1$ superconformal field theory.
\end{abstract}
\end{titlepage}


\renewcommand{\thefootnote}{\arabic{footnote}} \setcounter{footnote}{0}


\section{Introduction}
Perturbative renormalizability has been a criterion for the 
predictable quantum field theory.
Needless to say, this is because the renormalization removes ultraviolet (UV) divergences
of a Feynman diagram
and leads to a meaningful finite value of a physical quantity.
While an effective theory is permitted to include non-renormalizable interactions, of course,
this criterion must be satisfied 
by %
a fundamental theory without any cut-off scale
and excludes many models of the quantum field theory.
The renormalizable theories, however, do not exhaust all possibilities.

Actually a quantum theory endowed with a UV fixed point is well defined
and valid at the whole energy scale.
This possibility is known as the Weinberg asymptotic safety scenario
\cite{Weinberg:1979},
which perhaps rescues the non-renormalizability of the perturbative quantum gravity.
This scenario is also very attractive because
a renormalizable but asymptotically non-free theory such as pure QED
involves the Landau pole and the convergence radius of the perturbation becomes zero 
according to popular apprehension.
With assuming the existence of the UV fixed point,
we can avoid such a theoretical inconsistency included in perturbative quantum field theory.

UV fixed point is very important notion of the quantum field theory,
but it is very hard in general to determine whether a theory has a UV fixed point or not.
Five-dimensional minimal supersymmetric gauge theories are typical and attractive exceptions to this difficulty.
Perturbative five-dimensional gauge theories are non-renormalizable,
but Seiberg \cite{Seiberg:1996bd} showed that  perturbative description breaks down at high energy
but some of such theories flows up to a strongly coupling, non-Gaussian, UV fixed point.
$SU(2)$ gauge theory with $N_f=0,1,\cdots,7$ fundamental flavors
provides a concrete example \cite{Morrison:1996xf,Intriligator:1997pq} .
The flavor symmetry of this gauge theory is $SO(2N_f)\times U(1)_I$,
where $U(1)_I$ is associated with the instanton number.
The UV fixed point is described by a strongly coupled conformal field theory.
At this fixed point, the flavor symmetry is expected to enhance to
the larger group $E_{N_f+1}$:
$E_{1}=SU(2)$, $E_{2}=SU(2)\times U(1)$, $E_{3}=SU(3)\times SU(2)$,
$E_{4}=SU(5)$, $E_{5}=SO(10)$, and
$E_{6,7,8}$ are the usual exceptional Lie groups.

This enhancement of the flavor symmetries was conjectured by employing superstring theory
\cite{Seiberg:1996bd},
and so far it was not easy to show this enhancement based only on field theory arguments.
This is because the UV fixed point theories in question are strongly coupled,
and it has prevented us from verifying this conjecture directly.
Fortunately,
with recent progress in the theories of the localization and the superconformal index,
we can discuss the strongly-coupled fixed point theories quantitatively
by evaluating the protected indexes of these theories.
This direction was opened by Kim, Kim and Lee \cite{Kim:2012gu},
and their result was reformulated and extended
by the wok of Iqbal and Vafa \cite{Iqbal:2012xm}.
In this paper we study the detailed structure of their result on the five-dimensional
superconformal index, and we provide a justification of the enhancement of the flavor symmetry
for $N_f=0,1$.

The structure of the paper is as follows. 
We give a brief review on background materials in section 2. 
In section 3 we study some superconformal field theories associated with 5-brane web constructions
by computing the superconformal indexes of them. 
We conclude in section 4. 
In appendix A, we set some conventions, deriving useful formulas.

\section{5d Superconformal Index and Nekrasov formula}

In this paper we will discuss five-dimensional $\mathcal{N}=1$ superconformal field theories,
and these theories enjoy the superconformal symmetry named $F(4)$.
The bosonic part of $F(4)$
is $SO(2,5)\times SU(2)_R$.
The bosonic conformal group contains $SO(2)\times SO(5)$,
and this $SO(5)$ is the five-dimensional Lorentz group.
The $SO(2)$ charge gives the dimension $\epsilon_0$.
Let $A=1,2$ be the $SU(2)_R$ index and $m=1,2,3,4$ be the spinor index of $SO(5)$.
Let $Q_m^A$ and $S_A^m$ be the super and superconformal charges.

To define the superconformal index,
let us pick up the generators $\mathcal{Q}=Q_2^1$ and $\mathcal{S}=\mathcal{Q}^\dag=S_1^2$.
Then the above commutation realtion gives
\begin{align}
\Delta=
\{\mathcal{Q},\mathcal{S}\}=
\epsilon_0-2j_1-3R\geq 0.
\end{align}
The $SU(2)_L\times SU(2)_R\subset Sp(4)$ Cartans are $j_{1,2}$,
which are given by the Cartans of $SO(5)$ in the orthogonal basis $h_{1,2}$
\begin{align}
j_1=\frac{h_1+h_2}{2},\quad j_2=\frac{h_1-h_2}{2}.
\end{align}
The Cartan of $SU(2)_R$ is demoted by $R$.
 The states annihilated by $\mathcal{Q}$ and $\mathcal{S}$ are then
the $\frac{1}{8}$ BPS states.
Since they saturate the positivity bound,
these states have the dimension $\epsilon_0=2j_1+3R$.
To count these minimally BPS states we can use the Witten index
\begin{align}
I=\textrm{Tr}_{\mathcal{H}_{\frac{1}{8}\,\textrm{BPS}}}\,(-1)^F
=N_{\textrm{bosons}}-N_{\textrm{fermions}}.
\end{align}
This index counts the BPS states which are annihilated by  $\mathcal{Q}$ and $\mathcal{Q}^\dag=\mathcal{S}$.
Even in the case where $I$ is finite,
the numbers $N_{\textrm{bosons}}$ and $N_{\textrm{fermions}}$
are infinite in general.
To make the expression meaningful,
we need to introduce regulators 
which correspond to the Cartan generators commuting with $\mathcal{Q}$, $\mathcal{S}$, and each other.
In our case these commuting generators are
\begin{align}
\Delta, \quad
j_1+R,\quad
j_2,\quad
H_f,\quad
J=*\textrm{tr}(F\wedge F),
\end{align}
where $H_f$ are the Cartans of flavor symmetry,
and $J$ is the conserving instanton current.
We then arrive at the following definition of the so-called superconformal index:
\begin{align}
I_{\,\textrm{5d}}(u,z_f,t,q)
=\textrm{Tr}_{\mathcal{H}_{\frac{1}{8}\,\textrm{BPS}}}\,(-1)^F
e^{-\beta \{\mathcal{Q},\mathcal{S}\}}x^{2(j_1+R)}y^{2j_2}
\prod_fz_f^{H_f}u^k.
\end{align}
$k$ counts the instanton charge.
The fugacities $x$ and $y$ count the $j_{1,2}$ charges,
and we can introduce $SO(5)$ Cartan fugacities $t$ and $q$ through the relation
\begin{align}
x=\sqrt{\frac{q}{t}},\quad
y=\sqrt{qt}.
\end{align}

We can recast the index into the Euclidean path integral over fields with twisted boundary conditions
\begin{align}
I_{\,\textrm{5d}}(z_f,t,q)
=
\int_{\textrm{twisted b.c.}} \mathcal{D}\Phi \,e^{-S_{S^1\times S^4}[\Phi]}.
\end{align}
Kim, Kim, and Lee \cite{Kim:2012gu}
performed the path integration by employing the localization method\footnote{The localization on $S^1\times S^4$
was also formulated and computed in \cite{Terashima:2012ra,Nosaka:2013cpa}. }
 \textit{a la} Pestum,
and they found the following expression of the superconformal index:
\begin{align}
\label{KKLformula}
I_{\,\textrm{5d}}(z_f,t,q)
=\oint \prod_\alpha[{da_\alpha}]\vert Z_{\,\textrm{5d}}^{\,\textrm{Nek.}}(Q_\alpha,z_f;t,q)\vert^2,
\end{align}
where 
$[da_\alpha]$ is the Haar measure for loop variables $Q_\alpha=e^{ia_\alpha}$ and
we introduce new  fugacities as $q=xy$ and $t=\frac{y}{x}$.
Here $Z_{\,\textrm{5d}}^{\,\textrm{Nek.}}$ is the 5d Nekrasov partition function \cite{Nekrasov:2002qd},
and the integral is taken over $|Q_\alpha|=1$ for the loop variables $Q_\alpha=e^{ia_\alpha}$.
The conjugation $\overline{*}$ is defined by $\overline{f(Q_\alpha,z_f;t,q)}=f(Q_\alpha^{-1},z_f^{-1};t,q)$
for generic function.
The two integrants $Z_{\,\textrm{5d}}^{\,\textrm{Nek.}}$ and  $\overline{Z_{\,\textrm{5d}}^{\,\textrm{Nek.}}}$
are the contributions from the north and south poles of the sphere $S^4$,
where the fixed points of the localization computation.

Iqbal and Vafa \cite{Iqbal:2012xm} gave a string-theoretical reformulation of the above formula. Let us recall the idea of the geometric engineering:
a 5d $\mathcal{N}=1$ gauge theory arises from certain Calabi-Yau compactification of M theory.
As a consequence of the stringy realization,
the partition function of the 5d theory is exactly equal to the refined topological string partition function 
$Z_{\,\textrm{CY}}$ of the
corresponding CY.
This topological string partition function is defined by the refinement \cite{Awata:2005fa,Iqbal:2007ii} of the  topological vertex formalism
\cite{Aganagic:2002qg}-\cite{Bao:2011rc} ,
which is the large-$N$ dual to the refined Chern-Simons theory
through the geometric transition \cite{Gopakumar:1998ki,Aganagic:2012hs} .
See \cite{Awata:2005fa}-\cite{Iqbal:2012mt}  for details of this formalism.
The 5d index is then the following loop integral
\begin{align}
\label{IVF}
I_{\,\textrm{5d}}(z_f,t,q)
=\oint \prod_\alpha\frac{dQ_\alpha}{Q_\alpha}\,
Z_{\,\textrm{CY}}(Q_i;t,q)
Z_{\,\textrm{CY}}(Q_i^{-1};t^{-1},q^{-1})
,
\end{align}
where a loop variable $Q_\alpha$ is assigned for each loop in the web diagram,
and $Q=e^{-t}$ is the exponential of the K\"ahler parameter $t$. 
In fact, in this formula which was derived by Iqbal and Vafa
from the perspective of topological string theory,
the complex conjugation acts not only on $Q_i$ but also on $t$ and $q$,
and so the expression is slightly different from that of Kim, Kim and Lee.
In \cite{Iqbal:2012xm} the authors however
showed that usual topological string partition functions
satisfiy the relation
$Z_{\,\textrm{CY}}(Q_i^{-1};t^{-1},q^{-1})=Z_{\,\textrm{CY}}(Q_i^{-1};t,q)$,
and then we can recover the expression derived by Kim, Kim and Lee.
This is not always case
because when toric Calabi-Yau manifold contains a flat deformation direction of two cycle
as the case of $\mathcal{O}(0)\oplus \mathcal{O}(-2)\to\mathbb{CP}^2$,
the massive spectrum does not transforms as a representation of the Wigner little group \cite{Iqbal:2012xm, BMPTY,HKN}.
We therefore need to remove the factor coming from the problematic states which do not form any full spin content of
 $SU(2)_L\times SU(2)_R$.
The formalism to cleat the ugly contribution away was recently proposed in \cite{BMPTY,HKN},
and then we can obtain the renormalized partition function which enjoys the wanted invariance under the transformation
$t\to t^{-1}$ and $q\to q^{-1}$.

The point is that Iqbal and Vafa proposed that
this formula (\ref{IVF}) also works for 5d theories without any gauge theory description
nevertheless this Lagrangian description usually enables us to compute a Nekrasov partition function and 
the corresponding superconformal index exactly. 
This is because that
the refined topological vertex formalism \cite{Iqbal:2007ii} 
extended by Iqbal and Kozcaz \cite{Iqbal:2012mt} 
provides the topological string partition function
for any toric Calabi-Yau three-fold such as 
$\mathcal{O}(-3)\to\mathbb{P}^2$
which does not lead to any 5d gauge theory.
The resulting partition function then provides the 5d superconformal index 
via the Iqbal-Vafa proposal (\ref{IVF}).
In this way we can compute the 5d index
only by using the web diagram of a theory.

Actually, complete refined topological string partition functions 
including the constant map contribution 
take the following form\footnote{We omit the classical contribution
because it does not contribute to the superconformal index.}
 \cite{Iqbal:2012xm}:
\begin{align}
\label{RTSPF}
 \left( M(t,q)\,M(q,t) \right)^{\frac{\chi (\textrm{CY})}{4}  }
\,Z_{\,\textrm{CY}}'
(Q_i;t,q),
\end{align}
where $M(t,q)$ is a two-parameter extension of the MacMahon function,
and $Z_{\,\textrm{CY}}'
$ is the part that is given by
the refined topological vertex.
Since we study toric Calabi-Yau manifolds with single compact four-cycle in this paper,
let the Euler number $\chi (\textrm{CY})$ be $2$ in the following.
The definition of this MacMahon function can be found in the next section.
If the M-theory compactified on a Calabi-Yau manifold
leads to a five-dimensional gauge theory,
the corresponding refined topological vertex partition function can be decomposed in the following two parts;
the perturbative and instanton contribution:
\begin{align}
&Z_{\,\textrm{CY}}'
(Q_F,Q_B;t,q)
=\sum_{\vec{Y}}
Q_B^{\vert\vec{Y} \vert}\,Z_{\vec{Y}}(Q_F;t,q)
=Z_{\,\textrm{CY}}^{\,\textrm{pert.}}(Q_F;t,q)Z_{\,\textrm{CY}}^{\,\textrm{inst.}}(Q_F,Q_B;t,q),\nonumber\\
&
Z_{\,\textrm{CY}}^{\,\textrm{pert.}}(Q_F;t,q)=Z_{\vec{\varPhi}}(Q_F;t,q).
\end{align}
The Nekrasov expression of the partition function
$Z_{\,\textrm{CY}}
=\sum_{\vec{Y}}
Q_B^{\vert\vec{Y} \vert}\,Z_{\vec{Y}}$
naturally arises from the cut-and-glue process
for toric diagrams in the topological vertex formalism.
The full partition function is thus the following combination
of the perturbative and instanton contributions:
\begin{align}
 Z_{\,\textrm{CY}}\,(Q_i;t,q)
=\sqrt{M(t,q)\,M(q,t)}
Z_{\,\textrm{CY}}^{\,\textrm{pert.}}(Q_i;t,q)\,Z_{\,\textrm{CY}}^{\,\textrm{inst.}}(Q_i;t,q).
\end{align}
This expression corresponds to the formula of Kim, Kim and Lee \cite{Kim:2012gu},
\begin{align}
I_{\,\textrm{5d}}\,(z_f;t,q)
=\int [da_\alpha]\,
\textrm{PE}[f_{\textrm{mat.}}(x,y,Q_m,Q_{F\alpha})+f_{\textrm{vec.}}(x,y,Q_{F\alpha})]\,
\cdot\vert I^{\,\textrm{inst.}}\vert^2,
\end{align}
where the contributions from the two fixed points correspond to
the instanton partition function $Z_{\,\textrm{CY}}^{\,\textrm{inst.}}= I^{\,\textrm{inst.}}$
and the remaining part gives $|\sqrt{M(t,q)M(q,t)}Z_{\,\textrm{CY}}^{\,\textrm{pert.}}|^2dQ/Q$.

\begin{figure}[tbp]
 \begin{center}
  \includegraphics[width=140mm,clip]{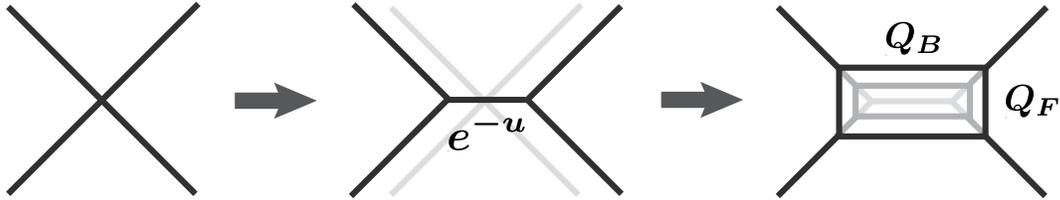}
 \end{center}
 \caption{The first deformation which moves the external lines is the mass deformation of the theory $E_1$.
 The second arrow denotes the breathing mode parametrized by $Q_F$.}
 \label{fig;def}
\end{figure}

The Coulomb branch deformation is parametrized by $Q_F$,
and masse deformations are
$Q_B$ and the K\"ahler parameters of the exceptional curves $Q_{Ei}$.
Interpretation of these parameters is as follows \cite{Aharony:1997ju,Aharony:1997bh}:
singular limit of a web diagram describes a superconformal point of the corresponding 5d theory.
The left hand side of \textbf{\textit{Fig}.\ref{fig;def} }illustrates the singular limit of the $SU(2)$ pure Yang-Mills theory.
This figure describes the web diagram of the toric Calabi-Yau three-fold
and the dual $(p,q)$ 5-brane web.
The deformation which moving the external lines  of the web diagram is a mass deformation
and parametrized by $u$ and $u_i$'s.
This global deformation moves infinitely massive branes,
and it is associated with the Cartan of the global symmetry of a 5d theory.
By giving VEV for the scalar component of the background vector multiplets for the global symmetry,
this deformation is generated.
This deformation thus changes the parameters of the theory.
On the one hand,
the breathing mode is a deformation of  the web without moving the external lines.
Only 5-branes with finite length are shifted under this deformation,
and this variation is the change of the size of the loops inside the web diagram.
We introduce the parameters $Q_F$ 
for each loop to parametrize this deformation\footnote{In this article
we focus only on toric Calabi-Yau manifolds with single loop.}.
By turning on the scalar component of the vector multiplets
of the local symmetry of a theory,
we can generate this deformation.
This is the deformation along the Coulomb branch of the theory.

In the following section,
we provide explicit equations which relate the K\"ahler parameters with the Coulomb branch parameters and the fugacities.


\section{Enhancement of Flavor Symmetry via 5d Index}
In this section,
we prove the enhancement of the flavor symmetry of $SU(2)$ gauge theories in the level of the
5d superconformal index.
In the pioneer work \cite{Kim:2012gu}
the enhancement was checked up to certain order of the power expansion of the index.
We will provide an all order proof for some $SU(2)$ gauge theories.

We introduce the fugacities $x$ and $y$ by the $\Omega$-background through the relation
\begin{align}
q=xy,\quad
t=\frac{y}{x},
\end{align}
and we expand a superconformal index as a positive power series in $x$.
To do so,
we have to introduce the following analytic continuation for combinatorial formulas.
For instance, the refined MacMahone function
\begin{align}
M(t,q)=\prod_{i,j=1}^\infty (1-q^it^{j-1})^{-1}=
\prod_{i,j=1}^\infty (1-x^{i-j+1}y^{i+j-1})^{-1},
\end{align}
appears in the superconformal indexes.
Since the naive geometric power series expansion of this expression
contains infinitely many negative power of $x$,
certain analytic continuation is needed to obtain the appropriate series expansion in $x\ll 1$.
We employ the continuation
in the sense of the relation $\sum_{n=0}^\infty t^n=-\sum_{n=0}^\infty t^{-n-1}$.
Then the preferred expression of MacMahone function in region $|x|\ll 1$ becomes\footnote{Since
the function $M(t,q)$ has dense poles along $\vert t\vert=1$,
this deformation of the formula is not proper analytic continuation.
We deal with a partition function as a formal power series in $t$ and $q$ with imposing the relation $\sum_{n\in\mathbb{Z}}t^n=\delta(t-1)$.
See \cite{MacDonnaldSymmetric} for rigorous treatment of such functions.}
\begin{align}
M(t,q)&=
\exp\left(\,
\sum_{n=1}^\infty \frac{1}{n}\frac{q^n}{(1-q^n)(1-t^n)}\right)\nonumber\\
&=
\exp\left(\,-
\sum_{n=1}^\infty \frac{1}{n}\frac{q^nt^{-n}}{(1-q^n)(1-t^{-n})}\right)
\nonumber\\
&\label{AC}
=\prod_{i,j=1}^\infty (1-q^it^{-j})=
\prod_{i,j=1}^\infty (1-x^{i+j}y^{i-j}).
\end{align}
This expression immediately implies a positive-power expansion with respect to $x$,
and therefore we employ such an continuation procedure to get the 
appropriate series expression
of a superconformal index.

The formula (\ref{IVF}) then gives the following expression 
\begin{align}
I_{\,\textrm{5d}}\,(u;t,q)
=M(t,q)\,M(q,t)\,
\oint \prod_\alpha\frac{dQ_\alpha}{2\pi i Q_\alpha}\,
Z_{\,\textrm{CY}}^{\,\textrm{t.v.}}(Q_i;t,q)
Z_{\,\textrm{CY}}^{\,\textrm{t.v.}}(Q_i^{-1};t,q)
,
\end{align}
where we use the identity $M(t^{-1},q^{-1})=M(q,t)$ to get this expression.
In the following we investigate the detailed structure of this combinatorial expression
in some concrete examples.
This formula
has the overall factor $M(t,q)M(q,t)$.
We need a converging positive power series expansion in a variable $|x|\ll 1$,
however analytic continuation to this region implies
\begin{align}
M(t,q)M(q,t)&=\exp \left(
\sum_{n=1}^\infty
\frac{1}{n}\frac{t^n+q^n}{(1-q^n)(1-t^n)}
\right)\nonumber\\
&=\exp \left(
\sum_{n=1}^\infty
\frac{1}{n}\frac{(xy)^n+(y/x)^n}{(1-(xy)^n)(1-(y/x)^n)}
\right)\nonumber\\
&=\exp \left(
-\sum_{n=1}^\infty
\frac{1}{n}\left(
\frac{x^n\left((y)^n+(1/y)^n\right)}{(1-(xy)^n)(1-(x/y)^n)}+1
\right)
\right),
\end{align}
and then the argument of the exponential function is crearly not converging.
We therefore have to introduce the following modified factor
by eliminating the diverging contribution $\sum_n1/n$
\begin{align}
\mathscr{M}(x,y)^2= {M(t,q)\,\tilde{M}(q,t)}
=\exp \left(
-\sum_{n=1}^\infty
\frac{1}{n}
\frac{x^n\left((y)^n+(1/y)^n\right)}{(1-(xy)^n)(1-(x/y)^n)}
\right).
\end{align}
Actually the same regularization is also employed in the paper of Kim, Kim and Lee
to derive the one loop partition function from the localization computation.

\subsection{$E_1$ and $\tilde{E}_1$ theories}

The $SU(2)$ gauge theories in 5d are special
because there exist two theories
distinguished by two topologically inequivalent configurations $\pi_4(SU(2))=\mathbb{Z}_2$ \cite{Douglas:1996xp}.
This discrete group $\mathbb{Z}_2$ labels the allowed two value of the theta angle $\vartheta$.
We start with the UV fixed point theory for the pure Yang-Mills theory with vanishing theta angle.
The local $\mathbb{F}_0$ is the canonical line bundle over 
the Hirzebruch surface $\mathbb{F}_0=\mathbb{P}^1\times\mathbb{P}^1$,
and this local Calabi-Yau threefold
engineers this five-dimensional minimal-supersymmetric $SU(2)$ 
Yang-Mills theory.
At the strong coupling, 
all the compact two cycles in the toric web diagram collapse
and the diagram becomes singular as the left hand side of \textbf{\textit{Fig}.\ref{fig;def}} .
The deformation from the SCFT point is
illustrated in \textbf{\textit{Fig}.\ref{fig;def}}.
The global mode $u$ is then given by $Q_B=uQ_F$,
and the local deformation is parametrized by $Q_F=e^{2ia}$.

\subsubsection*{$E_1$ theory and local $\mathbb{F}_0$ partition function}

By turning on a relevant operator deformation associated with
the background gauge field kinetic term for the Cartan sub-algebra of the flavor symmetry,
$E_1$ SCFT
flows down to 5d $SU(2)$ pure super Yang-Mills theory \cite{Seiberg:1996bd},
which is engineered by 
Type IIA superstring theory on
the local Hirzebruch surface $\mathbb{F}_0$.
A point is that we can compute the index of the UV SCFT by using this IR gauge theory
since the superconformal index is RG-invariant quantity.
The conventional topological string partition function \cite{Iqbal:2007ii},
namely the Nekrasov partition function \cite{Nekrasov:2002qd},
which provides this superconformal index 
is
\begin{align}
 Z_{\,\mathbb{F}_0}\,(Q_B,Q_F;t,q)
&=\mathscr{M}(x,y)\,
\rule{0pt}{5ex}
\sum_{\vec{R}}
Q_B^{|\vec{R}|}\,
q^{\parallel R_2^t\parallel^2}\, t^{\parallel R_1^t\parallel^2}\,
\widetilde{Z}_{R_1}(t,q)\,\widetilde{Z}_{R_2}(q,t)\,
\widetilde{Z}_{R_1^t}(q,t)\,\widetilde{Z}_{R_2^t}(t,q)\nonumber\\
&\quad\times
\prod_{i,j=1}^\infty
\frac{1}
{\left(1-Q_Ft^{i-1-R_{2,j}}q^{j-R_{1,i}} \right)
\left(1-Q_Ft^{j-R_{2,i}}q^{i-1-R_{1,j}} \right)}.
\end{align}
Here $\mathscr{M}(x,y)$ is the constant map contribution,
and $\widetilde{Z}_{R}$ is a normalized Macdonald function
in the principal specialization.
See Appendix.A for the concrete definition of these function.
The partition function can be factorized into the one-loop and instanton parts:
\begin{align}
Z_{\,\mathbb{F}_0}\,(Q_B,Q_F;t,q)=\mathscr{M}(x,y)\,Z^{N_f=0}_{\,\textrm{pert}}(Q_F;t,q)
\,Z^{N_f=0}_{\,\textrm{inst}}(Q_B,Q_F;t,q).
\end{align}
Here the combination $Z^{N_f=0}_{\,\textrm{pert}}Z^{N_f=0}_{\,\textrm{inst}}$
is the topological string partition function for the local $\mathbb{F}_0$
which is given by the refined topological vertex formalism.
The perturbative and instanton part of the five-dimensional Nekrasov partition function
take the following forms
\begin{align}
&Z^{N_f=0}_{\,\textrm{pert}}\,(Q_F;t,q)
=\prod_{i,j=1}^\infty(1-Q_Fq^it^{j-1})^{-1}(1-Q_Fq^{i-1}t^{j})^{-1},
\label{purepert}
\\
&Z^{N_f=0}_{\,\textrm{inst}}\,(uQ_F,Q_F;t,q)
=\sum_{\vec{Y}}
\left( u\frac{q}{t} \right)^{\vert \vec{Y}\vert}
\prod_{\alpha,\beta=1}^2\frac{1}{N_{Y_\alpha,Y_\beta}(Q_{\beta\alpha};t,q)}
,
\end{align}
where $u$ is the instanton factor, and the combinatorial factors are
\begin{align}
&N_{R,Y}(Q;t,q)=\prod_{(i,j)\in R}(1-Qt^{Y^t_j-i}  q^{R_i-j+1})
\prod_{(i,j)\in Y}(1-Qt^{-R^t_j+i-1}  q^{-Y_i+j}),\\
&Q_{21}=Q_F=Q_{12}^{-1},\quad Q_{11}=Q_{22}=1.
\end{align}
See appendix for details on the Nekrasov partition functions.
Let us consider the instanton expansion 
\begin{align}
Z_{\textrm{inst}}(uQ_F,Q_F;t,q)=1+\sum_{n=1}^\infty u^{n}Z^{(n)}(Q_F;t,q).
\end{align}
Then the each contribution with fixed instanton number satisfies the following equality:
\begin{align}
Z^{(n)}(Q_F^{-1};t^{-1},q^{-1})=Z^{(n)}(Q_F;t,q)=Z^{(n)}(Q_F^{-1};t,q)
\end{align}
We give a proof of this relation in Appendix.A.
Then the integrant of the superconformal index becomes 
\begin{align}
\vert Z^{N_f=0}_{\,\textrm{inst}}\,(uQ_F,Q_F;t,q)\vert^2
&=Z^{N_f=0}_{\,\textrm{inst}}\,((uQ_F)^{-1},Q_F^{-1};t,q)\,Z^{N_f=0}_{\,\textrm{inst}}\,(uQ_F,Q_F;t,q)\\
&=\sum_{n=0}^{\infty}(u^n+u^{-n})F^{(n)}(Q_F;x,y)
\end{align}
where the fugacities in \cite{Kim:2012gu} are given by $q=xy$ and $t=\frac{y}{x}$ and
$F^{(n)}$ is
\begin{align}
&F^{(0)}(Q;x,y)
=\frac{1}{2}\sum_{\ell=0}^\infty Z^{(\ell)}(Q;t,q)^2,\\
&F^{(n)}(Q;x,y)
=\sum_{\ell=0}^\infty Z^{(\ell)}(Q;t,q)Z^{(n+\ell)}(Q;t,q),\quad n\neq0,
\end{align}
for $Z^{(0)}=1$.
Then, this formula enables us to derive the nontrivial invariance of superconformal indexes under $u\to u^{-1}$,
and so we can rewrite the instanton contribution as a function of the $E_1=SU(2)$ characters
\begin{align}
\chi_{\bf 2s+1}^{E_1}(u)=u^s+u^{s-1}+\cdots +u^{-s+1}+u^{-s}.
\end{align}
Since the remaining factors inside the superconformal index are independent of $u$,
this formula gives a non-trivil evidence of the conjecture that the index is a function of $SU(2)$ characters. 
Since the index is not a generic function of $u$ but a function of $u^s+u^{-s}$,
it is suggested that the flavor symmetry is enhanced from $U(1)$ to $SU(2)$ in all instanton numbers.

\begin{figure}[tbp]
 \begin{center}
  \includegraphics[width=60mm,clip]{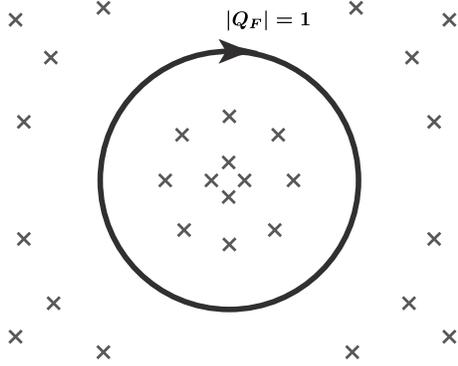}
 \end{center}
 \caption{In the limit $\vert x\vert \ll 1$ the pole region of the integrant of the formula (\ref{indexE0})
 splits into near the origin and faraway region.}
 \label{fig;poles}
\end{figure}

Using the defining equation (\ref{purepert}), we can easily show
\begin{align}
Z_{\textrm{pert}}(Q_F^{-1},t^{-1},q^{-1})
=Z_{\textrm{pert}}(Q_F^{-1},t,q),
\end{align}
and then we get the following expression
\begin{align}
 I_{\,\textrm{5d}}^{\,N_f=0}\,(u;x,y)
=&\mathscr{M}(x,y)^2\,
\sum_{n=0}^{\infty}(u^n+u^{-n})\nonumber\\
&\times\oint_{\vert Q\vert=1} \frac{dQ}{4\pi iQ}Z_{\textrm{pert}}(Q;x,y)Z_{\textrm{pert}}(Q^{-1};x,y)F^{(n)}(Q;x,y).
\label{indexE0}
\end{align}
To evaluate the index, we first expand this integrant into the positive-power series of $x$,
and then we collect the $Q$-independent part of it.
We can also compute the index by evaluating the contour integration directly.
As we will see below, all the poles in the integrant of our formula (\ref{indexE0}) are located inside or outside
of the unit circle $\vert Q\vert=1$ in the parameter region $\vert x\vert \ll 1$ as \textbf{\textit{Fig}.\ref{fig;poles}}.
We can therefore compute the residue integration by picking up all the
poles inside the unit circle,
and we can evaluate the integral without any regularization or ambiguity
 in $\vert x\vert \ll 1$.

Let us move on to explanation of how this formula works.
In $\vert x\vert\ll 1$ we have the \textit{unique} power series expansion of the perturbative
partition function in $Q_F$ and $x$
\begin{align}
Z_{\textrm{pert}}
&=\prod_{i,j}(1-Q_Fq^it^{j-1})^{-1}(1-Q_Fq^{i-1}t^{j})^{-1}\nonumber\\
&=\prod_{i,j}(1-Q_Fq^it^{-j})(1-Q_Fq^{i-1}t^{1-j})\nonumber\\
&=1-Q_F\bigg(1+\chi_2(y)x+(\chi_2(y))^2x^2+((\chi_2(y))^3-\chi_2(y))x^3+\cdots\bigg)\nonumber\\
&+Q_F^2\bigg(\chi_2(y)x+(1+(\chi_2(y))^2)x^2)+(2(\chi_2(y))^3-\chi_2(y))x^3+\cdots\bigg)\nonumber\\
&\rule{0pt}{4ex}-Q_F^3\big(x^2
+
(\chi_2(y))^3x^3+\cdots\big)+\cdots,
\end{align}
and therefore we obtain the power series expression of  the perturbative contribution to the index
$Z_{\textrm{pert}}(Q_F,t,q)Z_{\textrm{pert}}(Q_F^{-1},t,q)$.
The instanton contribution is more complicated, however, we can check that
power expansion in $|x|\ll 1$ implies the unique $Q_F$-series expression.
This is because a Nekrasov partition function is a summation over rational functions
and each pole structure of a rational function takes the form $(1-Q_F^lx^my^n)^{-1}$.
We therefore expand these factors into the following geometric power series
\begin{align}
&\frac{1}{1-Q_F^lx^my^n}=1+Q_F^lx^my^n+Q_F^{2l}x^{2m}y^{2n}+\cdots,\quad\textrm{if } m>0,\\
&\frac{1}{1-Q_F^lx^my^n}=-Q_F^{-l}x^{-m}y^{-n}-Q_F^{-2l}x^{-2m}y^{-2n}+\cdots\quad\textrm{if } m<0,
\end{align}
and then we obtain a positive power expansion in $x$.
However, if $m=0$ factor appeared, this algorithm would not not work because
this criterion for getting $x$ series does not implies any preferred series expansion
of $(1-Q_F^ly^n)^{-1}$.
Fortunately, some cancellation mechanism eliminates such seeming poles.
For instance, the one-instanton part of $SU(2)$ partition function is
\begin{align}
\frac{q}{t}u
\frac{1}{(1-q)(1-t^{-1})}\left(\,
\frac{1}{(1-Q_F^{-1})(1-Q_Ft^{-1}q)}
+\frac{1}{(1-Q_F)(1-Q_F^{-1}t^{-1}q)}
\,\right)
,
\end{align}
and these two fixed point contributions have a common problematic pole $(1-Q_F)^{-1}$.
But we can collect these two contributions into the following single rational functions 
\begin{align}
\frac{uQ_F(q+t)}{(1-q)(1-t)(Q_F-t/q)(Q_F-q/t)}.
\end{align}
Therefore one-instanton partition function does not contain any problematic pole
and we can obtain the unique series expansion.
This idea works also for higher instanrton numbers,
and the denominator of the two-instanton partition function for instance is proportional to
\begin{align}
(Q_F-t/q)(Q_F-t/q^2)(Q_F-t^2/q)(Q_F-q/t)(Q_F-q/t^2)(Q_F-q^2/t),
\end{align}
and there is no pole at $Q_F=(qt)^n$.
This fact also holds for the Yang-Mills theory with the Chern-Simons level $m=1,2$.
We can thus determine the expansion of the instanton contribution
in the variables $x$ and $Q_F$.
This series expression immediately implies the superconformal index
through our formula.

\subsubsection*{Pole structure and AGT relation}

As we have observed experimentally, 
there is no problematic pole $\left(1- Q_F(tq)^n\right)$
in the denominator of the instanton partition functions.
We can verify this fact rigorously using the AGT relation \cite{Alday:2009aq,Wyllard:2009hg}.
The AGT relation claims that the Nekrasov partition function
of a 4d $\mathcal{N}=2$ gauge theory is equal to the conformal block of the corresponding 2d CFT.
The instanton number is the conformal level in the 2d theory.
The denominator in the $k$-instanton partition function,
which is the level-$k$ part of the conformal block,
is the level-$k$ Kac determinant.
This determinant
\begin{align}
\det K_k
=c_k\prod_{1\leq rs\leq k}
(h-h_{r,s})^{P(k-rs)}
\end{align}
where $P$ is the number $\prod_{n=1}(1-q^n)^{-1}=\sum_{m=0}P(m)q^m$,
$h$ is the conformal dimension, and the zeros of the determinant is
\begin{align}
h_{r,s}=\frac{Q^2-(rb+s/b)^2}{4},\quad
Q=b+\frac{1}{b}.
\end{align}
We parametrize the central charge as $c=1+6Q^2$,
and the $\Omega$-background is related to the Liouville background charge
$b=\epsilon_1$ and $1/b=\epsilon_2$.

Let us apply this fact to our problem.
If the problematic pole $\left(1-Q_F(tq)^n\right)$ exists in the five-dimensional
partition function,
it implies the following pole in the denominator of the partition function in the
four-dimensional limit:
\begin{align}
\left(2a-n(b-1/b)\right)\left(2a+n(b-1/b)\right).
\end{align}
Notice the symmetry $n\to -n$ in the denominator.
The AGT relation is satisfied under the identification between 
the conformal dimension and the Coulomb branch parameter $h=Q^2/4-a^2$,
this problematic factor corresponds to 
\begin{align}
h-\frac{Q^2-(nb-n/b)^2}{4}.
\end{align}
Such a pole at $h_{n,-n}$ is obviously absent from the Kac determinant,
and therefore we conclude that the instanton partition function 
is free from any problematic poles.

Actually, our logic works only if the five-dimensional uplift
of the four-dimensional denominator is just the $q$-number extension 
$\det (K_k)
\propto \prod (2a-a_{rs})^{p_{rs}} \to \prod (1-e^{R(2a-a_{rs})})^{p_{rs}}$
of the Kac determinant.
This fact is verified for pure $SU(2)$ Yang-Mills theory
with zero Chern-Simons level $m_{\textrm{eff}}=0$.
The Kac determinant of the $q$-Virasoro algebra \cite{qVir,qW} is given by
\begin{align}
\det K_k
=\prod_{1\leq rs\leq k}\left(
\frac{(1-q^r)(1-t^r)}{q^r+r^r}
\right)^{P(k-rs)}
(\lambda^2-\lambda^2_{r,s})^{P(k-rs)},
\end{align}
where $\lambda_{r,s}=q^{-s/2}t^{r/2}+q^{s/2}t^{-r/2}$.
The AGT relation in five dimensions \cite{Awata:2009ur} suggests the following map
between the Coulomb branch parameter and the dimension of the
corresponding primary $\lambda$
\begin{align}
\lambda=Q_F^{\frac{1}{2}}+Q_F^{-\frac{1}{2}}.
\end{align}
The Kac determinant therefore takes the form 
\begin{align}
(\lambda^2-\lambda^2_{r,s})^{P(k-rs)}=
\left(Q_F-q^st^{-r}\right)^{P(k-rs)}
\left(1-{q^{-s}t^{r}}Q_F^{-1}\right)^{P(k-rs)},
\end{align}
and no factor in question  appears.
There is therefor no problematic pole in the denominators of Nekrasov partition functions
if the AGT relation is satisfied.
The idea and formulation toward the proof of the five-dimensional AGT relation  in the case of $N_c=2$, $N_f=0$ and $m=0$
is proposed in \cite{Yanagida:2010,Awata:2011ce}
.

\subsubsection*{$\tilde{E}_1$ theory and local $\mathbb{F}_1$ partition function}
Let us move on to
the computation of the index for the UV fixed point theory $\tilde{E}_1$
which arises from the $SU(2)$ Yang-Mills theory with non-vanishing theta term $\vartheta\equiv1\,\,\textrm{mod}2$.
It is straightforward to compute the superconformal index in this case.

The full Nekrasovpartition function is
\begin{align}
Z_{\,\mathbb{F}_1}\,(Q_B,Q_F;t,q)=\mathscr{M}(x,y)\,Z^{N_f=0}_{\,\textrm{pert}}(Q_F;t,q)
\,Z^{N_f=0}_{m=1
}(Q_B,Q_F;t,q),
\end{align}
where the instanton partition function is 
\begin{align}
Z^{N_f=0}_{m=1
}(Q_B,Q_F;t,q)&=
\sum_{\vec{Y}}
\left( -u\frac{q}{t} \right)^{\vert \vec{Y}\vert}
\frac{
 Q_F^{\frac{\vert Y_1\vert-\vert Y_2\vert}{2}}
\prod_\alpha t^{-\frac{\parallel Y^t_\alpha \parallel^2 }{2}}
q^{ \frac{\parallel Y_\alpha  \parallel^2  }{2}}
}{\prod_{\alpha,\beta=1}^2 N_{Y_\alpha,Y_\beta}(Q_{\beta\alpha};t,q)}=\sum_{n=0}^\infty
u^nZ^{(n)}_{m=1}(Q_F,t,q).
\end{align}
In this case
\begin{align}
{Z}^{(n)}_{m=1}\,(Q_F;t,q)
 =\sum_{\vert \vec{Y}\vert=n}
\frac{\left(- \frac{q}{t} \right)^{\vert \vec{Y}\vert}
 Q_F^{\frac{\vert Y_1\vert-\vert Y_2\vert}{2}}
\prod_{\alpha=1,2} t^{-\frac{\parallel Y^t_\alpha \parallel^2 }{2}}
q^{ \frac{\parallel Y_\alpha  \parallel^2  }{2}}
}{\prod_{\alpha,\beta=1}^2 N_{Y_\alpha,Y_\beta}(Q_{\beta\alpha};t,q)}
.
\end{align}
By applying this instanton partition function we can compute the superconformal index as the formula in the previous example.

\subsection{New 5d SCFT from local $\mathbb{F}_2$?}
\begin{figure}[tbp]
 \begin{center}
  \includegraphics[width=120mm,clip]{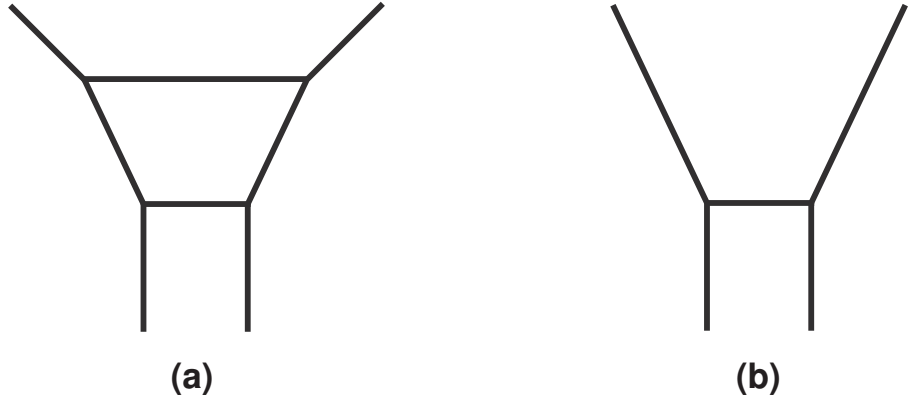}
 \end{center}
 \caption{The local $\mathbb{F}_2$ geometry and its six-dimensional contribution.
 (a) 5-brane web for the local $\mathbb{F}_2$ geometry. 
 (b) on the coincided two 5-branes, additional six-dimensional degrees of freedom appear. }
 \label{fig;F2}
\end{figure}
The Yang-Mills theories with effective Chern-Simons levels $m=0,1$
are engineered geometrically by the local Calabi-Yau threefold
over the Hirzebruch surfaces $\mathbb{F}_m$.
Actually these two theories corresponding to the two allowed values of the discrete theta angle $\vartheta=0,\pi$.
The UV fixed point theories of them are the known superconformal field theories: $E_1$ and $\tilde{E}_1$,
We have studied  the superconformal indexes of them.

There exists an another geometry in the sequence of the Hirzebruch surfaces, known as
$\mathbb{F}_2$.
The local geometry over this surface gives the Yang-Mills theory with $m=2$,
however no UV fixed point theory is known so far in the strong coupling limit.
One possibility is hence that this Yang-Mills theory is meaningless in the strongly-coupled UV
region, and it is only a cut-off theory
and needs the UV completion by adding additional degrees of freedom to make sense.
Actually the brane construction in the strong coupling limit provides additional massless
states
since the two parallel 5-branes coincides and the six-dimensional massless excitations
on them would couple to the boundary five-dimensional theory 
\textbf{\textit{Fig.}\ref{fig;F2}}.
Another possibility is that the six-dimensional degrees of freedom are actually decoupled
from the boundary five-dimensional theory,
and we get a new five-dimensional fixed point theory at the strong coupling limit.
We can not make a judgement only from the brane realization.
In the following we find the solution to the question by computing the superconformal index.

The Nekrasov partition function for $SU(2)$ pure Yang-Mills theory with the Chern-Simons level $2$,
which corresponds to the topological string partition function for the local $\mathbb{F}_2$ geometry,
takes the form
\begin{align}
&Z_{\,\mathbb{F}_2}\,(Q_B,Q_F;t,q)=\mathscr{M}(x,y)\,Z^{N_f=0}_{\,\textrm{pert}}(Q_F;t,q)
\,Z^{N_f=0}_{m=2}(Q_B,Q_F;t,q)\\
\rule{0pt}{6ex}&
Z^{N_f=0}_{m=2}(Q_B,Q_F;t,q)=\sum_{\vec{Y}}
\left( u\frac{q}{t} \right)^{\vert \vec{Y}\vert}
\frac{
 Q_F^{\vert Y_1\vert-\vert Y_2\vert}
\prod_\alpha t^{-{\parallel Y^t_\alpha \parallel^2 }}
q^{{\parallel Y_\alpha  \parallel^2  }}
}{\prod_{\alpha,\beta=1}^2 N_{Y_\alpha,Y_\beta}(Q_{\beta\alpha};t,q)}.
\end{align}
In order to use this formula for our index computation,
we introduce the relation between the physical parameters
\begin{align}
Q_F=e^{2ia},\quad
Q_B=u.
\end{align}
This Nekrasov partition function then
implies the following superconformal index through the formula (\ref{KKLformula})
\begin{align}
I_{\,\mathbb{F}_2}\,(u;x,y)
=&1+x^2+2\chi_2(y)x^3
+\left(2\chi_3(y)+1\right)x^4
 \nonumber\\
&+\bigg\{
-\left(u+\frac{1}{u}\right)(\chi_3(y)-1 ) -1+\chi_3(y)+2\chi_4(y)
\bigg\}x^5
 \nonumber\\
&+
\bigg\{
-\left(u+\frac{1}{u}\right)(\chi_3(y)+2 ) -1+3\chi_3(y) 
+2\chi_5(y)
\bigg\}x^6\nonumber\\
&+
\bigg\{
-\left(u+\frac{1}{u}\right)(2\chi_2(y)+\chi_4(y))
+\chi_2(y)+4\chi_4(y)+2\chi_6(y)
\bigg\} x^7
+\nonumber\\
&+\bigg\{\,
6+6\chi_3(y)+7\chi_5(y)+2\chi_7(y)
\nonumber\\
&
-\left(u+\frac{1}{u} \right)\left(
-1+3\chi_3(y)+\chi_5(y)
\right)
+\left(u^2+\frac{1}{u^2} \right)
 \bigg\}x^8
+\cdots.
\end{align}
Many coefficients of the terms in this expanded index are negative.
In general, the negative terms of an index correspond to the fermonic contributions.
But the above expression seems to be unusual because all the suerconformal indexes computed in \cite{Kim:2012gu}
have only a few negative terms. 
In the above case, we can save this unnaturalness technically just by redefining the fugacity $u$
as 
\begin{align}
u \rightarrow -u.
\end{align}
We however find a more natural solution to this problem.
Our claim is that the above expression contains certain extra contribution
in addition to that from 5d fixed point theory.
In the following we divide an unnecessary contribution
from the index,
and we then obtain the ``proper'' five-dimensional superconformal index whose
almost all the coefficients are positive.

Then, what is this extra contribution to the superconformal index?
In the 5-brane construction \textit{\textbf{Fig}}\textbf{.\ref{fig;F2}},
there exist two parallel 5-branes
which coincide with each other at the superconformal fixed point.
This fact suggests that the theory coming from a 5-brane web can contain
some massless degrees of freedom on the stack of the coincided 5-branes.
This has been pointed out in \cite{BMPTY, HKN}
by studying the five-dimensional uplift of the Gaitto $T_N$ theory.
We can also identify 
the extra contribution which makes original index non-proper  as
the six-dimensional degrees of freedom on the 5-brane stack.
In the language of the Calabi-Yau compactification,
this stack of parallel 5-branes is dual to the local structure $\mathcal{O}(0)\oplus
\mathcal{O}(-2)\to\mathbb{CP}^1$
in the local $\mathbb{F}_2$ geometry \textit{\textbf{Fig}}\textbf{.\ref{fig;F2}(b)}.
The moduli space of the BPS states of the compctification onto this geometry,
that is the moduli space of M2-branes wrapping the two cycles,
is then non-compact because of the flat $\mathcal{O}(0)$ direction,
and the resulting BPS states do not form full spin content\footnote{The author 
thanks his collaborators on the resent paper \cite{BMPTY}:
L.Bao, V.Mitev, E.Pomoni and
F.Yagi. 
He also thanks C.Koz\c{c}az for discussion on this problem  
through the collaborators.
See also the references \cite{Iqbal:2007ii, Iqbal:2012xm}.}.
To eliminate the extra contribution caused by this non-compactness,
we factor the index (or partition function) of the local $\mathcal{O}(0)\oplus
\mathcal{O}(-2)\to\mathbb{CP}^1$ geometry out of the index (or partition function).
This procedure is a simplified version of that for $N_f\geq 2$ theories found in \cite{BMPTY, HKN},
and this
extra factor coming from the  non-full spin content
is a five-dimensional analogue of the propotional coefficient of the AGT relation 
between Liouville correlators and the partition functions of gauge theories in four dimensions.

Let us move on to the computation of this extra prefactor.
Since this non-full spin content contribution is the partition function on the local geometry
\textit{\textbf{Fig}}\textbf{.\ref{fig;F2}(b)},
the refined topological vertex formalism provides the explicit expression.
We can therefor get the closed expression for the partition function of this non-full spin content
by using the refined topological vertex formalism
\begin{align}
Z_{\mathcal{O}(0,-2)_{\mathbb{CP}^1}}&=Z_{\textrm{non-full}}(u;x,y)\nonumber\\
&\rule{0pt}{4ex}=\sum_Y C_{\emptyset\emptyset Y}(q,t)(-Q_B)^{\vert Y\vert}f_Y(q,t)
C_{\emptyset\emptyset Y^t}(t,q)
\nonumber\\
&=
\prod_{i,j=1}^\infty\frac{1}{(1-ut^{i-1}q^{j})}.
\end{align}
In order to get the last equality we apply the formula (\ref{Macdonaldsum}).
We employ the following connected expression to get positive power expansion in $x$
\begin{align}
Z_{\,\textrm{non-full}}(u;x,y)
=\prod_{i,j=1}^\infty{(1-ut^{-i}q^{j})}=
\prod_{i,j=1}^\infty{(1-ux^{i+j}y^{-i+j})}.
\end{align}
The proper instanton contribution to the 5d theory is then given by  the following normalized partition function
\begin{align}
Z_{\,m=2}^{\textrm{full spin}}(u,&Q;x,y)=
\frac{Z_{\,m=2}^{\,N_f=0}(u,Q;x,y)}
{Z_{\textrm{non-full}}(u;x,y)}
=1+ux^2+u\chi_2(y)x^3
+\bigg(u\bigg(Q+1+\frac{1}{Q}+\chi_3(y)\bigg)
\nonumber\\
&+u^2\bigg)x^4+\left(u\chi_2(y)\left(Q+1+\frac{1}{Q}\right)+u\chi_4(y)
+u^2\chi_2(y)\right)x^5+
\cdots.
\end{align}
Then the normalized superconformal index is
\begin{align}
\tilde{I}_{\,\mathbb{F}_2}\,(u;x,y)
=\mathscr{M}(x,y)\,
\oint \frac{dQ_F}{4\pi i Q_F}\,
&\big| Z_{\,\textrm{pert}}^{\,{N_f=0}}(Q_F;x,y)\big|^2\nonumber\\
&\times Z_{\,m=2}^{\,\textrm{full spin}}(u,Q_F;x,y)
Z_{\,m=2}^{\,\textrm{full spin}}(u^{-1},Q_F^{-1};x,y)
.
\end{align}
Notice that the perturbative contribution
is not affected by the  ``Chern-Simons level" $m=2$.
By computing the residue integral, we find
\begin{align}
&\tilde{I}_{\,\mathbb{F}_2}\,(u;x,y)\nonumber\\
&=
1+\chi_{\bf 3}^{E_1}(u)x^2+\chi_2(y)\left(1+\chi_{\bf 3}^{E_1}(u)\right)x^3
+\bigg\{\, \chi_3(y)\left(1+\chi_{\bf 3}^{E_1}(u)\right)+1+\chi_{\bf 5}^{E_1}(u)\bigg\}x^4\nonumber\\
&+\bigg\{\, 
\chi_4(y)\left(1+\chi_{\bf 3}^{E_1}(u)\right)
+\chi_2(y)\left(1+\chi_{\bf 3}^{E_1}(u)+\chi_{\bf 5}^{E_1}(u)\right)
\bigg\}x^5\nonumber\\
&+\bigg\{\, 
\chi_5(y)\left(1+\chi_{\bf 3}^{E_1}(u)\right)
+2\chi_3(y)\left(1+\chi_{\bf 3}^{E_1}(u)+\chi_{\bf 5}^{E_1}(u)\right)
-1+\chi_{\bf 3}^{E_1}(u)+\chi_{\bf 7}^{E_1}(u)
\bigg\}x^6\nonumber\\
&+\bigg\{\, 
\chi_6(y)\left(1+\chi_{\bf 3}^{E_1}(u)\right)
+\chi_4(y)\left(2+4\chi_{\bf 3}^{E_1}(u)+2\chi_{\bf 5}^{E_1}(u)
\right)
\nonumber\\
&+\chi_3(y)\left(
1+3\chi_{\bf 3}^{E_1}(u)+2\chi_{\bf 5}^{E_1}(u)+\chi_{\bf 7}^{E_1}(u)
\right)
\bigg\}x^7+
\bigg\{\, 
\chi_7(y)\left(1+\chi_{\bf 3}^{E_1}(u)\right)\nonumber\\
&+\chi_5(y)\left(4+5\chi_{\bf 3}^{E_1}(u)+3\chi_{\bf 5}^{E_1}(u)\right)
+\chi_3(y)\left(2+7\chi_{\bf 3}^{E_1}(u)+3\chi_{\bf 5}^{E_1}(u)+2\chi_{\bf 7}^{E_1}(u)\right)\nonumber\\
&+
3+2\chi_{\bf 3}^{E_1}(u)+2\chi_{\bf 5}^{E_1}(u)+\chi_{\bf 9}^{E_1}(u)\bigg\}x^8
+
\bigg\{\, 
\chi_8(y)\left(1+\chi_{\bf 3}^{E_1}(u)\right)\nonumber\\
&+
\chi_6(y)\left(4+7\chi_{\bf 3}^{E_1}(u)+3\chi_{\bf 5}^{E_1}(u)\right)+
\chi_4(y)\left(6+10\chi_{\bf 3}^{E_1}(u)+6\chi_{\bf 5}^{E_1}(u)+3\chi_{\bf 7}^{E_1}(u)\right)\nonumber\\
&+
\chi_2(y)\left(4+7\chi_{\bf 3}^{E_1}(u)+4\chi_{\bf 5}^{E_1}(u)+2\chi_{\bf 7}^{E_1}(u)+\chi_{\bf 9}^{E_1}(u)\right)\bigg\}x^9
+\cdots.
\end{align}
This result is precisely equal to
that of $E_1$ SCFT ${I}_{\,\mathbb{F}_0}\,(u;x,y)$ computed in \cite{Kim:2012gu,Iqbal:2012xm},
and so we can conclude that the 5d fixed point theory
associated with local $\mathbb{F}_2$ is 
precisely the well known $E_1$ SCFT
arising from $SU(2)$ pure Yang-Mills theory without discrete theta term $m=0$\footnote{This fact 
has been conjectured in \cite{Bergman:2013ala} very recently
based on the one-instanton partition function,
and our result provides a multi-instanton justification of this conjecture.}.
This result is natural from the perspective of Calabi-Yau compactification.
The surface $\mathbb{F}_2$ is a complex structure deformation of the surface $\mathbb{F}_0$,
and the topological string partition function, namely the Nekrasov partition function,
is independent of such a deformation.
The partition function of $\mathbb{F}_2$ is then equal to that of $\mathbb{F}_0$ if this independence 
on the complex structure is satisfied.
The refined topological string partition function however can jump  
under the complex structure deformation,
and the surfaces $\mathbb{F}_{0,2}$ are different in the Nekrasov partition function.
This difference is, however, very simple as we found
\begin{align}
\label{wallcross}
Z_{\,\textrm{c.s.d.of }\mathbb{F}_0}
=Z_{\,\textrm{non-full }}\cdot
Z_{\,\mathbb{F}_0},
\end{align}
where $Z_{\,\textrm{c.s.d.of }\mathbb{F}_0}$ is the partition function of the complex structure deformation of $\mathbb{F}_0$,
namely $\mathbb{F}_2$.
The jump of the refined Gopakumar-Vafa invariants 
therefore can be collected into the overall factor
$Z_{\,\textrm{non-full }}$ which is the non-full spin content of the $\mathbb{F}_2$ theory.
This structure is very similar to that of the wall crossing phenomena,
and it would be very interesting if we can justify our expectation (\ref{wallcross})
based on the theory of the refined Gopakumar-Vafa invariants and its wall crossing.

Moreover, our result gives an evidence for the conjecture that
the contribution from non-compact parallel 5-branes
can be eliminated by factoring the partition function
of the non-full spin content out.
This conjecture is originally based on the comparison between 
$SU(2)$ $2\leq N_f \leq 5$ gauge theory and the expected $E_{N_f+1}$ UV fixed point.
Our argument in this article is based on the 
relation between the discrete theta terms and the two known fixed point theories $E_1$ and $\tilde{E}_1$.
It therefore provides an another evidence of the above mentioned conjecture on the 
contribution of the non-full spin content.

Our result  suggests that the correspondence between generic Hirzebruch surface and UV fixed point theories.
For $m\geq3$, the Hirzebruch surface $\mathbb{F}_m$ is not Fano,
and is specified by a non-convex fan.
We can expect that the local geometry over $\mathbb{F}_m$ is related to the $E_s$ theory as $m\equiv s \mod 2$.
By computing the partition function for  $\mathbb{F}_m$ and its non-full spin content
along the line of \cite{BMPTY, HKN},
we can expect
\begin{align}
I_{\,\mathbb{F}_s}\,(u;x,y)
=\frac{I_{\,\mathbb{F}_{2N+s}}(u;x,y)}
{I^{\,\textrm{non-full}}_{\,\mathbb{F}_{2N+s}}(u;x,y)},\quad
s=0,1,\,\,N=0,1,2,3,\cdots.
\end{align}
We leave the detailed  check of this relation to future study.

\begin{figure}[tbp]
 \begin{center}
  \includegraphics[width=100mm,clip]{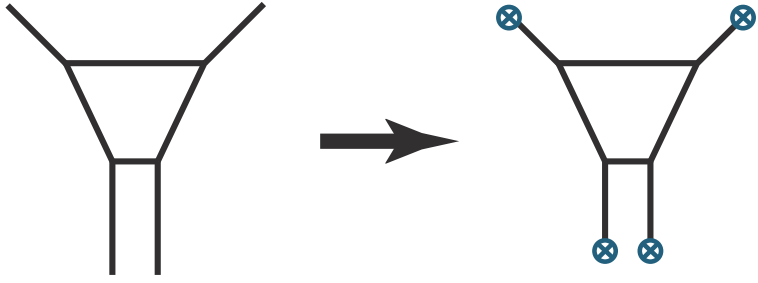}
 \end{center}
 \caption{We can regularize the 5-brane web by terminating
the external branes on 7-branes. 
The 7-branes are illustrated by the blue circles.
We conjecture that the regularized theory
 can be computed by eliminating the contribution from the non-full-spin content,
and the resulting 5d theory is
equal to the  local $\mathbb{F}_0$ theory, i.e. the $E_1$ SCFT. }
 \label{fig;F2and7}
\end{figure}

\subsection{$E_2$ theory}
The local second del Pezzo surface $dP_2$ 
engineers $SU(2)$ gauge theory with single fundamental flavor,
and it was shown that the topological string partition function for the geometry
is exactly equal to the five-dimensional Nekrasov partition function for the gauge theory.
The full Nekrasov partition function is
\begin{align}
&Z^{N_f=1}_{\,\textrm{pert}}\,(Q_F,z;t,q)
=\mathscr{M}(x,y)\,\prod_{i,j=1}^\infty(1-Q_Fq^it^{j-1})^{-1}(1-Q_Fq^{i-1}t^{j})^{-1}\nonumber\\
&\qquad\qquad\qquad\qquad\qquad\qquad\qquad\quad\times
(1-Q_Et^{i-\frac{1}{2}}q^{j-\frac{1}{2}})(1-Q_FQ_Et^{i-\frac{1}{2}}q^{j-\frac{1}{2}}t)
\\
&
Z^{N_f=1}_{\,\textrm{inst}}\,(uQ_F,Q_F,{z};t,q)\rule{0pt}{5ex}
=\sum_{\vec{Y}}
\left( u\frac{q}{t} \right)^{\vert \vec{Y}\vert}
\frac{
Z^{\vec{Y}}_{\textrm{fund}}(Q_a,Q_m={z};t,q)
}{
Z^{\vec{Y}}_{\textrm{vec}}(Q_a;t,q)}
,
\end{align}
where $Q_E=Q_m/{\sqrt{Q_F}}=z/{\sqrt{Q_F}}$.
The vector multiplet contribution takes the form
\begin{align}
Z^{\vec{Y}}_{\textrm{vec}}(Q_a;t,q)
=\frac{\left(\frac{q}{t}\right)^{\frac{N_c\vert\vec{Y}\vert}{2}}}{\prod_{\alpha,\beta}N^{\vec{Y}}_{\alpha,\beta}(Q_{\beta\alpha};t,q)}.
\end{align}
We can therefore show the following symmetry of this factor
\begin{align}
Z^{{(Y_1,Y_2)}}_{\textrm{vec}}(Q_a^{-1};t,q)
=Z^{{(Y_2,Y_1)}}_{\textrm{vec}}(Q_a;t,q),\quad
Z^{(Y_1,Y_2)}_{\textrm{fund}}(Q_a^{-1},Q_m;t,q)
=Z^{(Y_2,Y_1)}_{\textrm{fund}}(Q_a,Q_m;t,q).
\end{align}
This property leads to the relation
$Z^{N_f=1}_{\,\textrm{inst}}\,(uQ_F^{-1},Q_F^{-1},{z};t,q)=Z^{N_f=1}_{\,\textrm{inst}}\,(uQ_F,Q_F,{z};t,q)$,
and it implies the enhancement $U(1)\to SU(2)$ of the flavor symmetry associated with the fugacity $u$ as the case of pure Yang-Mills theory.

We can show an expression of the superconformal index very similar to (\ref{indexE0}).
The only difference is that the partition function in the expression is not that of pure Yang-Mills theory but that of $N_f=1$ theory in this case.
The parametrization is
$Q_B=ue^{2ia}$, $Q_F=e^{2ia}$ and $Q_E={z}e^{-ia}$ \cite{Iqbal:2012xm},
and so the integration is taken over $Q_F$.

\subsection{$E_3$ theory}

Let us consider $SU(2)$ gauge theory with two fundamental hypermultiplets.
The Nekrasov partition function is
\begin{align}
Z^{N_f=2}_{\,\textrm{inst}}\,(uQ_F,Q_F,{z}_1,{z}_2;t,q)
=\sum_{\vec{Y}}
\left( u\frac{q}{t} \right)^{\vert \vec{Y}\vert}
\frac{
\prod_{f=1,2}Z^{\vec{Y}}_{\textrm{fund}}(Q_a,Q_{mf};t,q)
}{\prod_{\alpha,\beta=1}^2 N_{Y_\alpha,Y_\beta}(Q_{\beta\alpha};t,q)}
,
\end{align}
where we use the following parametrization
\begin{align}
Q_{m1}={z}_1{z}_2,\quad
Q_{m2}=\frac{{z}_2}{{z}_1}.
\end{align}
It is easy to show the following invariance which suggests the enhancement  $U(1)\to SU(2)$
\begin{align}
Z^{N_f=2}_{\,\textrm{inst}}\,(uQ_F^{-1},Q_F^{-1},{z}_1,{z}_2;t,q)
=Z^{N_f=2}_{\,\textrm{inst}}\,(uQ_F,Q_F,{z}_1,{z}_2;t,q).
\end{align}

Since the partition function is invariant under the exchange of the masses $Q_{m1}\leftrightarrow Q_{m2}$,
we find the invariance
\begin{align}
Z^{N_f=2}_{\,\textrm{inst}}\,(uQ_F,Q_F,{z}_1^{-1},{z}_2;t,q)
=Z^{N_f=2}_{\,\textrm{inst}}\,(uQ_F,Q_F,{z}_1,{z}_2;t,q).
\end{align}
This suggests that the abelian flavor symmetry associated with ${z}_1=\sqrt{Q_{m1}/Q_{m2}}$
is also enhanced to $SU(2)$.
The following enhancement is therefore manifest in our formula,
\begin{align}
U(1)_{u}\times U(1)_{{z}_1}\times U(1)_{{z}_2}
\to
SU(2)\times SU(2)\times U(1)
\subset SU(2)\times SU(3)=E_3.
\end{align}
To get the full enhancement,
we have to eliminate the extra contribution as the case of $\mathbb{F}_2$ geometry.
See \cite{BMPTY,HKN} for details on this issue.


\section{Discussion}
In this article we have discussed on the enhanced flavor symmetry of five-dimensional UV fixed point theories
through the superconformal index.
We gave the expression of superconformal indexes in which the dependence on the special combination of instanton fugacity 
$u^s+u^{-s}$ becomes manifest,
and then we can find that the index is a function of the $SU(2)$ characters $u^s+u^{s-1}+\cdots +u^{-s}$. 
We extended our argument for some cases of higher-rank flavor symmetry
by using the structure of the Nekrasov partition function of SQCD.

We also compute the superconformal index associated with the local $\mathbb{F}_2$ geometry.
It had been known that
the local Hirzebruch surfaces $\mathbb{F}_0$ and $\mathbb{F}_1$
correspond to the two known superconformal field theories
with rank one flavor symmetry $E_1$ and $\tilde{E}_1$.
There exist, however, another Calabi-Yau manifolds and the corresponding 5-brane web configurations which lead to 
rank one flavor symmetry and one-dimensional Coulomb branch.
The local second Hirzebruch surface $\mathbb{F}_2$ is a typical example,
and it had been an open problem what is the superconformal field theory which appears at the UV fixed point of the $\mathbb{F}_2$ theory.
In this paper we pointed out that the five-dimensional theory arising from M-theory on the local $\mathbb{F}_2$ geometry
contains some extra degrees of freedom in the nature of an six-dimensioal excitation. 
We proposed the method to remove the extra contribution from the Nekrasov partition function,
and then found that the resulting superconfromal index is identical with that of  the $\mathbb{F}_0$ theory.
Therefore the superconformal field theory for $\mathbb{F}_2$ geometry should be precisely the $E_1$ theory
which is the UV fixed point theory of $SU(2)$ pure Yang-Mills theory with vanishing discrete theta term.
This fact was pointed out  very recently in \cite{Bergman:2013ala} based on the one instanton partition function,
and our argument is based on the multi-instanton computation and the superconfromal index.

It would be interesting to extend  our idea precisely to
$E_{N_f+1}$ theories for $N_f=2,\cdots,7$.
Another interesting application of our result would be the research on the unknown relation between UV fixed point theories 
associated with various 5-brane web configurations.
Since we can extract the proper superconformal index 
from the topological vertex partition function by removing the extra contribution which does not form full spin content,
we can check the relation between fixed point theories by comparing quantitatively the superconfromal indexes.
We leave these problems for future study.

\section*{Acknowledgments}
We are very grateful to Ling Bao, Can Koz{\c c}az, Vladimir Mitev, Elli Pomoni and
Futoshi Yagi for helpful discussions and suggestions during the investigation of the 5d $E_6$ SCFT.


\section*{Appendix A : Building blocks of the Nekrasov partition function}

\subsection*{A-1 Young diagram}

The arm and leg length are defined
\begin{align}
a_Y(i,j)=Y_i-j,\quad \ell_Y(i,j)=Y^t_j-i,\quad
a'(i,j)=j-1,\quad,
\ell'(ij)=i-1.
\end{align}
We introduce the norms of a Young diagram as
$\vert Y\vert:=\sum_{i=1}^{d(Y)}Y_i$ and
$\parallel Y\parallel^2:=\sum_{i=1}^{d(Y)}Y_i^2$.
\begin{align}
&n(Y)=\sum_{s\in Y} \ell_Y(s)=\sum_{s\in Y}\ell'(s)=\frac{\parallel Y^t\parallel^2-\vert Y\vert}{2}
\nonumber\\
&\kappa_Y=2\left( n(Y^t)-n(Y)\right).
\end{align}

\subsection*{A-2 : symmetric functions}
The refined MacMahon function is
\begin{align}
M(t,q)=\prod_{i,j=1}^\infty (1-q^it^{j-1})^{-1}.
\end{align}
The following specialized Macdonald function play a role in the refined topological vertex formalism:
\begin{align}
\widetilde{Z}_R(t,q)=\prod_{i=1}^{d(R)}\prod_{j=1}^{R_i} (1-q^{R_i-j}t^{R^t_j-i+1})^{-1}.
\end{align}

The Macdonald functions are basis of the ring of the symmetric functions over the field of rational functions $\mathbb{Q}(q,t)$.
We can define the Macdonald functions uniquely as the functions
\begin{align}
P_\lambda(x;q,t)
=
m_\lambda(x)
+\sum_{\mu\prec\lambda}
u_{\lambda\mu}m_\mu(x),\quad u_{\lambda\mu}\in\mathbb{Q}(t,q),
\end{align}
which satisfy the orthogonality 
\begin{align}
\langle P_\mu,P_\nu\rangle_{q,t}=0\quad \textrm{if}\quad
\mu\neq\nu.
\end{align}
The inner product is defined by using that for the power-sum symmetric functions
\begin{align}
\langle p_\mu,p_\nu\rangle_{q,t}
=\delta_{\mu\nu}z_\lambda\prod_{i=1}^{\lambda^t_1}\frac{1-q^{\lambda_i}}{1-t^{\lambda_i}}.
\end{align}

The norm of the Macdonald function is
\begin{align}
\langle P_\lambda,P_\lambda\rangle_{q,t}=\prod_{s\in \lambda}
\frac{1-q^{a(s)+1}t^{\ell(s)}}{1-q^{a(s)}t^{\ell(s)+1}}\equiv
\left(\frac{q}{t}\right)^{\frac{\vert \lambda\vert}{2}}\frac{1}{b_\lambda(q,t)}.
\end{align}
The Macdonald function satisfies the following Cauchy fornulas
\begin{align}
\label{Cauchy1}
&\sum_\lambda b_\lambda(q,t)P_\lambda(x;q,t)P_\lambda(y;q,t)=
\exp\left(
\sum_{n=1}^\infty
\frac{1}{n}
\left(\frac{q}{t}\right)^{\frac{n}{2}}
\frac{1-t^n}{1-q^n}p_n(x)p_n(y)
\right)
,\\
\label{Cauchy2}
&\sum_\lambda P_\lambda(x;q,t)P_{\lambda^t}(y;t,q)=\prod_{i,j}(1+x_iy_j).
\end{align}
These relations play the key role to take a summation over Young diagrams
and get a closed expression.

The principal specialization of the Macdonald function is important to study the Nekrasov formulas
\begin{align}
&P_\lambda(t^\rho;q,t)=
\prod_{s\in\lambda}
\frac{-{t}^{\frac{1}{2}}q^{a(s)}}{1-q^{a(s)}t^{\ell(s)+1}},\\
&P_{\lambda^t}(q^\rho;t,q)=
\prod_{s\in\lambda}
\frac{q^{-a(s)-\frac{1}{2}}}{1-q^{-a(s)-1}t^{-\ell(s)}}.
\end{align}
This specialization is related to $\widetilde{Z}_\lambda$ as
\begin{align}
\widetilde{Z}_\lambda(t,q)=
\left(-\sqrt{\frac{q}{{t}}}\right)^{\vert\lambda\vert}q^{-\frac{\parallel \lambda\parallel^2}{2}}
P_\lambda(t^\rho;q,t).
\end{align}

By using the Cauchy formula (\ref{Cauchy1}) for $x=y=t^\rho$,
we can obtain
\begin{align}
\label{Macdonaldsum}
\sum_\lambda
\left(
Q\sqrt{\frac{t}{{q}}}
\right)^{\vert\lambda\vert}q^{\parallel \lambda\parallel^2}\widetilde{Z}_\lambda(t,q)\widetilde{Z}_{\lambda^t}(q,t)
=\prod_{i,j=1}^\infty\frac{1}{1-Qt^{i-\frac{1}{2}}q^{j-\frac{1}{2}}}.
\end{align}

\subsection*{A-3 : Nekrasov partition function}

\subsubsection*{Pure Yang-Mills theory}

Let us introduce the following factor 
\begin{align}
N^{\vec{Y}}_{\alpha,\beta}(Q_{\beta\alpha};t,q)
&=
\prod_{s\in Y_\alpha}
\left( 1-Q_{\beta\alpha}\,t^{\ell_{Y_\beta}(s)}q^{a_{Y_\alpha}(s)+1} \right)
\prod_{t\in Y_\beta}
\left( 1-Q_{\beta\alpha}\,t^{-(\ell_{Y_\alpha}(t)+1)}q^{-a_{Y_\beta}(t)} \right)\nonumber\\
&=
\prod_{i,j=1}^\infty
\frac{1-Q_{\beta\alpha}\,t^{-Y^t_{\alpha,j}+i-1}q^{-Y_{\beta,i}+j}}{1-Q_{\beta\alpha}\,t^{i-1}q^{j}}.
\end{align}
We also indicate the same function by the symbol $N_{Y_\alpha,Y_\beta}(Q_{\beta\alpha};t,q)$.
The vector multiplet contribution to the Nekrasov partition functions is,
\begin{align}
Z_{\,\textrm{vec}}^{\,\vec{Y}}=
\frac{\left(\frac{q}{t}\right)^{\frac{N_c\vert\vec{Y}\vert}{2}}}{\prod_{\alpha,\beta}N^{\vec{Y}}_{\alpha,\beta}(Q_{\beta\alpha};t,q)},
\end{align}
and the Chern-Simons term with the effective level $m_{\textrm{eff}}$ contributes in the following form
\begin{align}
Z_{\,\textrm{CS}}^{\,\vec{Y}, m_{\textrm{eff}}}=
\left(\frac{q}{t}\right)^{\frac{m_{\textrm{eff}}\vert\vec{Y}\vert}{2}}
\prod_\alpha Q_\alpha^{-m_{\textrm{eff}}\vert Y_\alpha\vert} \prod_{(i,j)\in Y_\alpha} 
t^{-m_{\textrm{eff}}(i-1)}q^{m_{\textrm{eff}}(j-1)}.
\end{align}

Then the five-dimensional Nekrasov partition function for $\mathcal{N}=2$ $U(N_c)$ pure SYM theory with the Chern-Simons level $m=m_{\textrm{eff}}$
is
\begin{align}
Z_{\,\textrm{5d}}^{\,m}(\vec{a},\Lambda,R;\epsilon_1,\epsilon_2)
&=\sum_{\vec{Y}}\,
{\mathfrak{Q}
}^{\vert\vec{Y}\vert}\,
\left(\frac{q}{t}\right)^{\frac{(N_c+m)\vert\vec{Y}\vert}{2}}
\frac{\prod_\alpha Q_\alpha^{-m\vert Y_\alpha\vert} \prod_{(i,j)\in Y_\alpha} 
t^{-m(i-1)}q^{m(j-1)}}
{\prod_{\alpha,\beta}N^{\vec{Y}}_{\alpha,\beta}(Q_{\beta\alpha};t,q)}\nonumber\\
&=
\sum_{\vec{Y}}\,
\mathfrak{Q}^{\vert\vec{Y}\vert}\,
\left(\frac{q}{t}\right)^{\frac{N_c\vert\vec{Y}\vert }{2} }
\frac{\prod_\alpha Q_\alpha^{-m\vert Y_\alpha\vert}
t^{-\frac{m\parallel Y^t_\alpha \parallel^2 }{2}}
q^{ \frac{m\parallel Y_\alpha  \parallel^2  }{2}}}
{\prod_{\alpha,\beta}N^{\vec{Y}}_{\alpha,\beta}(Q_{\beta\alpha};t,q)}
.
\end{align}
We use the instanton factor $\mathfrak{Q}=(R\Lambda)^{2N_c}$ as the weight to count the instanton number, which is five-dimensional analogue 
of the dimensional transmutation.

\subsubsection*{Symmetry of the $SU(2)$ vector multiplet contribution}
Let us consider the following $n$-instanton partition function of $SU(2)$ Yang-Mills theory
\begin{align}
Z^{(n)}(Q,t,q)=\sum_{\vert \vec{Y}\vert=n}
Z_{\,\textrm{vec}}^{\,\vec{Y}}(Q_{21}=Q).
\end{align}
This partition function enjoys the following symmetries
\begin{align}
Z^{(n)}(Q^{-1},t,q)=Z^{(n)}(Q,t,q),\quad
Z^{(n)}(Q,t^{-1},q^{-1})=Z^{(n)}(Q,t,q).
\end{align}
The first relation follows from the following relation
\begin{align}
Z_{\,\textrm{vec}}^{\,({Y_1},Y_2)}(Q_{21}=Q^{-1})
&=\left(\frac{q}{t}\right)^{\vert \vec{Y}\vert}
\frac{1}{N_{Y_1Y_1}(1)N_{Y_1Y_2}(Q)N_{Y_2Y_1}(Q^{-1})N_{Y_2Y_2}(1)}\nonumber\\
&=Z_{\,\textrm{vec}}^{\,({Y_2},Y_1)}(Q_{21}=Q).
\end{align}

The identity given in \cite{Taki:2007dh} 
\begin{align}
\sum_{s\in Y_\alpha}\ell_{Y_\beta}(s)
=\sum_{j=1}^{Y^t_{\alpha,1}}
\sum_{i=1}^{Y^t_{\alpha,j}}Y^t_{\beta,j}-\sum_{s\in Y_\alpha}i
=\sum_{j=1}^\infty
Y^t_{\alpha,j}Y^t_{\beta,j}-\frac{\parallel Y^t_\alpha\parallel^2}{2}-\frac{\vert Y_\alpha\vert}{2},
\end{align}
leads to the following relation
\begin{align}
N_{Y_1 Y_2}(Q;t^{-1},q^{-1})=(-Q)^{\vert \vec{Y}\vert}
t^{\frac{\parallel Y^t_1\parallel^2-\parallel Y^t_2\parallel^2+\vert Y_1\vert+\vert Y_2\vert}{2}}
q^{\frac{-\parallel Y_1\parallel^2+\parallel Y_2\parallel^2-\vert Y_1\vert-\vert Y_2\vert}{2}}
N_{Y_1 Y_2}(Q^{-1};t,q).
\end{align}
This identity immediately implies the relation $Z^{(n)}(Q,t^{-1},q^{-1})=Z^{(n)}(Q,t,q)$.

\subsubsection*{SQCD with fundamental flavors}

The fundamental hypermultiplet gives the following additional contribution: 
\begin{align}
Z^{\vec{Y}}_{\textrm{fund}}(Q_{\alpha},Q_m;t,q)
=\prod_\alpha\prod_{(i,j)\in Y_\alpha}
2\sinh
\frac{R
\left(
a_\alpha+m+\epsilon_1\left({j}-\frac{1}{2}\right)
\epsilon_2\left({i}-\frac{1}{2}\right)
\right)}{2}
,
\end{align}
To compare it with topological string partition functions,
we introduce the following combinatorial factor
\begin{align}
H^{\vec{Y}}
(Q_{\alpha},Q_m;t,q)
=\prod_\alpha\prod_{(i,j)\in Y_\alpha}
\left(
1-Q_m^{-1} Q_\alpha^{-1}t^{-i+\frac{1}{2}}q^{j-\frac{1}{2}}\right)
=N^{({Y_1},\emptyset)}_{12}\left(
Q_m^{-1} Q_\alpha^{-1}\sqrt{\frac{t}{q}}
;t,q
\right)
,
\end{align}
where $Q_\alpha=e^{-Ra_\alpha}$ and  $Q_m=e^{-Rm}$.
We then find
\begin{align}
Z^{\vec{Y}}_{\textrm{fund}}(Q_{\alpha},Q_m;t,q)
=(-\sqrt{Q_m})^{\vert\vec{Y}\vert}
Z_{\,\textrm{CS}}^{\,\vec{Y}, m=-\frac{1}{2}}
H^{\vec{Y}}(Q_{\alpha},Q_m;t,q),
\end{align}
where the Chern-Simons contribution is given by
\begin{align}
Z_{\,\textrm{CS}}^{\,\vec{Y}, m}
=\prod_\alpha Q_\alpha^{-m\vert Y_\alpha\vert}
t^{-\frac{m\parallel Y^t_\alpha \parallel^2 }{2}}
q^{ \frac{m\parallel Y_\alpha  \parallel^2  }{2}}.
\end{align}

\subsubsection*{Formula}

\begin{align}
Z_{\,\textrm{vec}}^{\,\vec{Y}}
&=
(-1)^{N \vert\vec{Y}\vert}\,
\mathfrak{C}_{\vec{Y}}(Q_\alpha)
\prod_\alpha \,
t^{\frac{(2\alpha-N)\parallel Y^t_\alpha \parallel^2 }{2}}
q^{ \frac{(N+2-2\alpha)\parallel Y_\alpha  \parallel^2  }{2}}
\tilde{Z}_{Y_\alpha}(t,q)
\tilde{Z}_{Y_\alpha^t}(q,t)\nonumber\\
&\times
\prod_{1\leq \alpha <\beta\leq N}
\prod_{i,j=1}^\infty
\frac{(1-Q_{\beta\alpha}q^{j}t^{i-1})(1-Q_{\beta\alpha}q^{j-1}t^{i})}
{
(1-Q_{\beta\alpha}q^{-Y_{\beta,i}+j}t^{-Y_{\alpha,j}^t+i-1})
(1-Q_{\beta\alpha}q^{-Y_{\beta,i}+j-1}t^{-Y_{\alpha,j}^t+i})
}
,
\end{align}

\begin{align}
\mathfrak{C}_{\vec{Y}}(Q_\alpha)=
\prod_{\alpha=1}^{N-1}
\left(Q_{\alpha,\alpha+1}\right)^{\,\,
-(N-\alpha)\sum_{\beta=1}^{\alpha}\vert Y_\beta\vert
-\alpha\sum_{\beta=\alpha+1}^N\vert Y_\beta\vert
}.
\end{align}

\renewcommand{\theequation}{A.\arabic{equation}}\setcounter{equation}{0}
\renewcommand{\thesubsection}{A.\arabic{subsection}}\setcounter{subsection}{0}



\end{document}